\documentclass[twocolumn,epjc3]{svjour3}

\usepackage[utf8]{inputenc}
\usepackage{microtype}
\usepackage{mathptmx}
\usepackage{amsmath}
\usepackage{amssymb}
\usepackage{xcolor}
\usepackage{tikz}

\usetikzlibrary{external}
\title{Non-quadratic improved Hessian PDF reweighting and application to CMS dijet measurements at 5.02 TeV}

\journalname{Eur. Phys. J. C}

\author{Kari J.\ Eskola\thanksref{addr1,addr2,e1} \and Petja Paakkinen\thanksref{addr1,addr2,e2} \and Hannu Paukkunen\thanksref{addr1,addr2,e3}}
\institute{University of Jyvaskyla, Department of Physics, P.O. Box 35, FI-40014 University of Jyvaskyla, Finland \label{addr1} \and Helsinki Institute of Physics, P.O. Box 64, FI-00014 University of Helsinki, Finland \label{addr2}}
\thankstext{e1}{e-mail: kari.eskola@jyu.fi}
\thankstext{e2}{e-mail: petja.paakkinen@jyu.fi}
\thankstext{e3}{e-mail: hannu.paukkunen@jyu.fi}

\begin{document}

\maketitle

\begin{abstract}
	Hessian PDF reweighting, or ``profiling'', has become a widely used way to study the impact of a new data set on parton distribution functions (PDFs) with Hessian error sets. The available implementations of this method have resorted to a perfectly quadratic approximation of the initial $\chi^2$ function before inclusion of the new data. We demonstrate how one can take into account the first non-quadratic components of the original fit in the reweighting, provided that the necessary information is available. We then apply this method to the CMS measurement of dijet pseudorapidity spectra in proton--proton (pp) and proton--lead (pPb) collisions at 5.02 TeV. The measured pp dijet spectra disagree with next-to-leading order (NLO) theory calculations using the CT14 NLO PDFs, but upon reweighting the CT14 PDFs, these can be brought to a much better agreement. We show that the needed proton-PDF modifications also have a significant impact on the predictions for the pPb dijet distributions. Taking the ratio of the individual spectra, the proton-PDF uncertainties effectively cancel, giving a clean probe of the PDF nuclear modifications. We show that these data can be used to further constrain the EPPS16 nuclear PDFs and strongly support gluon nuclear shadowing at small $x$ and antishadowing at around $x \approx 0.1$.
\end{abstract}

\section{Introduction}

The proton structure at high momentum-transfer, as encoded in the collinearly factorized parton distribution functions (PDFs), is not only an interesting subject in its own right, but plays a pivotal role in many applications, such as precision electroweak and Higgs physics, searches for new physics, etc.~\cite{Gao:2017yyd}. Likewise, their counterparts for nucleons bound in nuclei, the nuclear PDFs (nPDFs), are essential in e.g.~studying the production of hard probes of the Quark Gluon Plasma~\cite{Mangano:2004wq}. In practice, despite the ongoing effort in lattice methods~\cite{Lin:2017snn}, the PDFs are obtained by the well-established means of global analysis using hard-process data. As such, the PDFs have uncertainties which derive from those in the available data and also from the lack of data constraints in certain phase-space regions. It is then often the case that when new data are published or a future experiment is being planned, one would like to study the impact that the measurement could have on the PDFs. A good example of such a case is the recent CMS measurement of dijet pseudorapidity spectra in proton--proton (pp) and proton--lead (pPb) collisions at 5.02 TeV~\cite{Sirunyan:2018qel}, where, on one hand, the measured pp spectra seem to be in a disagreement with next-to-leading order (NLO) perturbative QCD (pQCD) calculations using CT14~\cite{Dulat:2015mca} and MMHT14~\cite{Harland-Lang:2014zoa} PDFs (see the Supplemental Material of Ref.~\cite{Sirunyan:2018qel}), while, on the other hand, the nuclear-modification ratio of the pPb and pp spectra appear to have much smaller uncertainties than predictions with various nPDFs. One should therefore study the impact these data could have on both the free-proton PDFs and their nuclear modifications.

As producing a full global fit remains rather involved, even with publicly available tools like the xFitter~\cite{Bertone:2017tig} (built upon the former HERAFitter~\cite{Alekhin:2014irh}) coming available, it is in most cases impractical for a general user to try to learn about the constraining power of a data set in this way. For this purpose, approximative methods have been developed, first in the formalism of Bayesian reweighting of Monte Carlo PDF ensembles~\cite{Giele:1998gw,Ball:2010gb,Ball:2011gg,Watt:2012tq,Sato:2013ika} and later in a framework using Hessian error sets~\cite{Paukkunen:2013grz,Paukkunen:2014zia,Schmidt:2018hvu}. These methods have their limitations, as the new PDFs rely on all the theoretical assumptions of the original PDF analysis, such as the parametrization form, the value of $\alpha_{\rm s}$ and the used heavy-quark scheme. There are also limitations related to how well the methods approximate the true parameter likelihood in the region constrained by the new data. In particular, the applications of Hessian PDF reweighting have resorted to a perfectly quadratic approximation of the $\chi^2$ goodness-of-fit function before inclusion of the new data and to linear or up to quadratic terms for responses in the new observables. This applies to the implementation in the xFitter package, where the method is referred to as ``Hessian profiling'', as well as to the new software package which has appeared under the name ePump~\cite{Schmidt:2018hvu}. It is not, however, uncommon that non-quadratic terms in the $\chi^2$ function are large (see e.g.~Figure~6 of Ref.~\cite{Martin:2009iq}), and thus it would be beneficial to have a way to take these into account.

The purpose of this article is twofold: First, in Section~\ref{sec:hessianrw}, we describe how one can include into Hessian PDF reweighting the first non-quadratic terms in the $\chi^2$ function consistently with the original fit, provided that the needed information is available. Second, in Section~\ref{sec:dijets}, we apply the Hessian PDF reweighting to the aforementioned CMS dijet measurements at 5.02 TeV~\cite{Sirunyan:2018qel}. We show that the strong disagreement between the pp measurement and next-to-leading order (NLO) calculations using CT14 NLO PDFs~\cite{Dulat:2015mca} can be brought to a much better agreement upon reweighting the CT14 PDFs, but that this requires rather strong modifications for high-$x$ gluons. We demonstrate that such changes in the proton PDFs have also an important impact on predictions for dijet production in pPb. Finally, we then reweight the EPPS16 nPDFs~\cite{Eskola:2016oht} with the nuclear modification ratio of the measured pPb and pp dijet spectra using the non-quadratic approximation developed in Section~\ref{sec:hessianrw} and present a discussion on the importance of these higher-order terms in the reweighting. Preliminary work on this topic can be found in Refs.~\cite{Eskola:2018xjr,Eskola:2018sxu}.

\section{PDF uncertainties and reweighting in Hessian method} \label{sec:hessianrw}

In this section, we first recapitulate the uncertainty determination in the Hessian approach~\cite{Pumplin:2001ct}, assuming the use of a global tolerance criterion. We then describe how one can perform a reweighting upon such determined error sets, taking into account the first non-quadratic terms in the $\chi^2$ function. We end the section with a discussion on the applicability of this method in the case of non-global tolerances.

\subsection{Hessian uncertainties with global tolerance criterion} \label{sec:hessianuncrt}

In PDF global analyses, the goodness-of-fit of a parameter vector $\vec{a}$ is dictated by the $\chi^2$ function
\begin{equation}
	\chi^2(\vec{a}) = \sum_{ij}\,(y_i(\vec{a}) - y_i^\text{data})\,C^{-1}_{ij}\,(y_j(\vec{a}) - y_j^\text{data}), \label{eq:chisq}
\end{equation}
where $y_i(\vec{a})$ are theory predictions for the observables included in the analysis, $y_i^\text{data}$ the corresponding measured values and $C^{-1}_{ij}$ the elements of the inverse covariance matrix for these data. In the Hessian method for uncertainty estimation, one takes the parameter values $\vec{a}^\text{min}$ which minimize Equation~\eqref{eq:chisq}, $\chi^2(\vec{a}^\text{min}) \equiv \min\chi^2(\vec{a}) \equiv \chi^2_0$, as the central, best-fit values and studies the behaviour of the $\chi^2$ function around this minimum to determine the uncertainty in these parameters.

The leading deviations from the minimum value $\chi^2_0$ are given by the quadratic approximation
\begin{equation}
  \chi^2 \approx \chi^2_0 + \sum_{ij}\,(a_i - a_i^\text{min})\,H_{ij}\,(a_j - a_j^\text{min}), \label{eq:chisqexp}
\end{equation}
where $H_{ij} = \frac{1}{2} \partial^2 \chi^2 / \partial a_i \partial a_j|_{\vec{a}=\vec{a}^\text{min}}$ are the elements of the Hessian matrix. In practice, these elements need to be obtained numerically. Since the Hessian matrix is symmetric, it has a complete set of orthonormal eigenvectors $\vec{v}^{(k)}$ such that
\begin{gather}
  \sum_j\,H_{ij}\,v^{(k)}_j = \epsilon_k\,v^{(k)}_i, \\
  \sum_i\,v^{(k)}_i\,v^{(\ell)}_i = \delta_{k\ell}, \qquad \sum_k\,v^{(k)}_i\,v^{(k)}_j = \delta_{ij},
\end{gather}
where $\epsilon_k$ are the eigenvalues of the Hessian matrix. With this eigendecomposition we can define new parameters
\begin{equation}
  z_k = \sum_i \sqrt{\epsilon_k}\,v^{(k)}_i\,(a_i - a_i^\text{min})
\end{equation}
such that Equation~\eqref{eq:chisqexp} becomes
\begin{equation}
  \chi^2 \approx \chi^2_0 + \sum_{k}\,z_k^2. \label{eq:chisqquadrappr}
\end{equation}

Since the new parameters $z_k$ are uncorrelated in the quad\-ra\-tic approximation, one can use the standard law of error propagation to translate the uncertainties in the parameters $z_k$ to the uncertainty of any PDF-dependent quantity $X$ as~\cite{Pumplin:2001ct}
\begin{equation}
  \Delta X = \sqrt{\sum_k \left(\frac{\partial X}{\partial z_k} \Delta z_k\right)^2}.
\end{equation}
Given a well justified global tolerance $\Delta\chi^2$ for the allowed growth of $\chi^2$ from its minimum, one can determine the allowed parameter variations $\Delta z_k$.\footnote{The intricacies of choosing an appropriate value for $\Delta\chi^2$ are outside the scope of this article, see Refs.~\cite{Martin:2009iq,Pumplin:2001ct,Pumplin:2009bb} for discussion.} If the $\chi^2$ function were perfectly quadratic, the uncertainty of the parameter $z_k$ corresponding to the tolerance $\Delta\chi^2$ would be simply $\Delta z_k = \sqrt{\Delta\chi^2}$. As this is generally not true, one instead finds $\delta z^\pm_k$, the positive and negative values of $z_k$ corresponding to the $\Delta\chi^2$ increase, and assigns $\Delta z_k = (\delta z^+_k - \delta z^-_k) / 2$. It is convenient to define error sets $S^\pm_i$ corresponding to parameter values
\begin{equation}
	z_k[S^\pm_i] =
	\begin{cases}
		\delta z^\pm_i, & k = i \\
		0, & k \neq i
	\end{cases}
	\quad, \label{eq:errorsets}
\end{equation}
along with the central set $S_0$, where $z_k[S_0] = 0$ for all $k$. Estimating
\begin{equation}
	\frac{\partial X}{\partial z_k} = \frac{X[S^+_k] - X[S^-_k]}{2\,\Delta z_k},
\end{equation}
where $X[S^\pm_k]$ stands for the quantity $X$ calculated with the parameter set of Equation~\eqref{eq:errorsets}, yields then a simple form
\begin{equation}
  \Delta X = \frac{1}{2}\sqrt{\sum_k \left(X[S^+_k] - X[S^-_k]\right)^2}.
	\label{eq:sympresc}
\end{equation}
As the response in $X$ to the upward and downward parameter shifts can be uneven, one can alternatively specify an upward--downward asymmetric error prescription e.g.~with~\cite{Lai:2010vv}
\begin{equation}
	\delta X^\pm = \sqrt{\sum_k \left[ \substack{\max \\ \min} \left\{ X[S^+_k]-X[S_0],X[S^-_k]-X[S_0],0 \right\} \right]^2}.
	\label{eq:asympresc}
\end{equation}

\subsection{Non-quadratic reweighting} \label{sec:reweighting}

\begin{figure}[tb]
	\centering
	\begin{tikzpicture}[samples=100, scale=0.825]
		\draw[->] (-4.5,0) -- (4.5,0) node[above] {$z_k$};
		\draw[->] (0,-1) -- (0,6) node[right] {$\chi^2 - \chi^2_0$};
		\draw[dotted] (4.5,4.5) -- (-4.5,4.5) node[above] {$\Delta\chi^2$};
		\node[above] at (4.5,4.5) {\phantom{$\Delta\chi^2$}};
		\draw[semithick, dash pattern=on 5pt off 5pt, dash phase=3pt, color=black!10!red, domain=-3.25:3.25] plot (\x,\x*\x/2);
		\draw[semithick, dash pattern=on 3pt off 7pt, dash phase=-3pt, blue, domain=-3.25:3.25] plot (\x,\x*\x/2);
		\draw[dotted, color=black!10!red] (3,4.5) -- (3,0) node[below] {$\sqrt{\Delta\chi^2}\phantom{\Delta\chi}$};
		\draw[semithick, domain=-3:4.25] plot (\x,0.957355*\x*\x/2 - 0.084628*\x*\x*\x/2);
		\draw[dotted] (-2.75,4.5) -- (-2.75,0) node[below] {$\delta z_k^-$};
		\draw[dotted] (3.75,4.5) -- (3.75,0) node[below] {$\delta z_k^+$};
		\draw[fill=black] (-2.75,4.5) circle (0.06cm);
		\draw[fill=black] (3.75,4.5) circle (0.06cm);
	\end{tikzpicture}
	\\
	\begin{tikzpicture}[samples=100, scale=0.825]
	  \draw[->] (-4.5,0) -- (4.5,0) node[above] {$z_k$};
		\node[above] at (-4.5,0) {\phantom{$z_k$}};
	  \draw[->] (0,-3.5) -- (0,3.5) node[right] {$y_i - y_i[S_0]$};
	  \draw[semithick, dash pattern=on 5pt off 5pt, dash phase=3pt, color=black!10!red, domain=-3.5:3.5] plot (\x,0.83333*\x);
	  \draw[dotted, color=black!10!red] (3,2.5) -- (0,2.5) node[left] {$\frac{y_i[S_k^+] - y_i[S_k^-]}{2}$};
	  \draw[dotted, color=black!10!red] (3,2.5) -- (3,0) node[below] {$\sqrt{\Delta\chi^2}\phantom{\Delta\chi}$};
	  \draw[dotted, red] (-3,-3) -- (-3,0);
	  \draw[semithick, dash pattern=on 3pt off 7pt, dash phase=4pt, blue, domain=-3.25:4] plot (\x,0.833*\x - 0.0556*\x*\x);
	  \draw[dotted] (-3,-3) -- (0,-3) node[right] {$y_i[S_k^-] - y_i[S_0]$};
	  \draw[dotted] (3.75,2) -- (0,2) node[left] {$y_i[S_k^+] - y_i[S_0]$};
	  \draw[semithick, domain=-3:4.25] plot (\x,0.855*\x - 0.0858*\x*\x);
	  \draw[dotted] (-2.75,-3) -- (-2.75,0) node[above] {$\delta z_k^-$};
	  \draw[dotted] (3.75,2) -- (3.75,0) node[below] {$\delta z_k^+$};
	  \draw[fill=black] (-2.75,-3) circle (0.06cm);
	  \draw[fill=black] (3.75,2) circle (0.06cm);
	\end{tikzpicture}
	\caption{An illustration for the response of $\chi^2$ (top) and $y_i$ (bottom) with respect to a change of parameter $z_k$ in quadratic--linear (red, long dashed), quadratic--quadratic (blue, short dashed) and cubic--quadratic (black, solid) approximations.}
	\label{fig:nonquadrappr}
\end{figure}
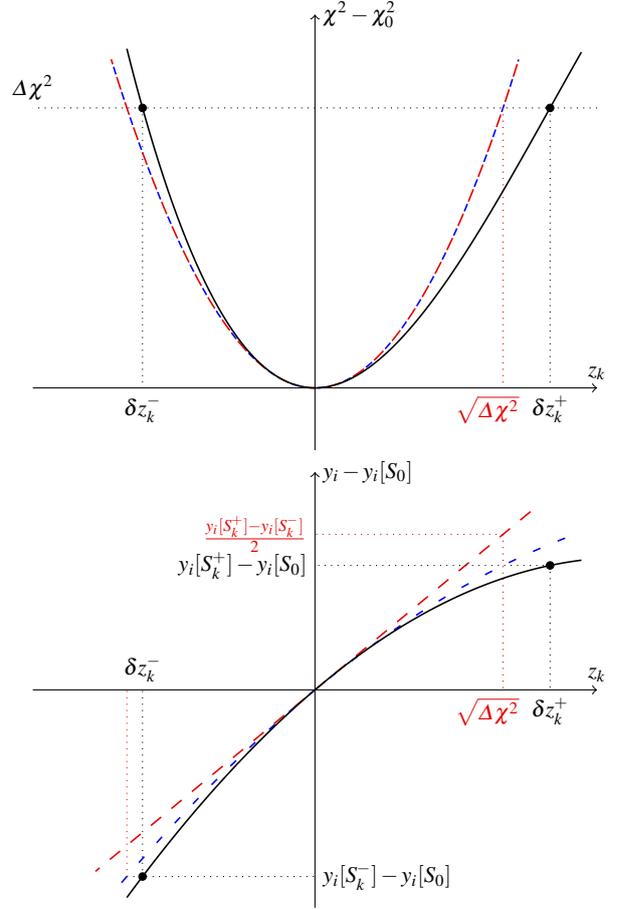

In the presence of a new data set, the total $\chi^2$ can be written as
\begin{equation}
	\chi^2_\text{new}(\vec{z}) = \chi^2_\text{old}(\vec{z}) + \sum_{ij}\,(y_i(\vec{z}) - y_i^\text{data})\,C^{-1}_{ij}\,(y_j(\vec{z}) - y_j^\text{data}), \label{eq:chisqnew}
\end{equation}
where $y_i$ ($y_i^\text{data}$) now correspond to the new theoretical (measured) values and $\chi^2_\text{old}$ incorporates our knowledge of the original global analysis. Now, as we do not wish to produce a full global analysis with $\chi^2_\text{new}$, we need to make suitable approximations. The simplest choice is to use the quadratic approximation in Equation~\eqref{eq:chisqquadrappr}, according to the method introduced in Ref.~\cite{Paukkunen:2014zia}, but if the parameter variations $\delta z^\pm_k$ and the global tolerance $\Delta\chi^2$ of this fit are known (as is the case with EPPS16 nPDFs, see Table~2 in Ref.~\cite{Eskola:2016oht}), then $\chi^2_\text{old}$ can be approximated with a third order polynomial in each of the eigendirections,
\begin{equation}
	\chi^2_\text{old} \approx \chi^2_0 + \sum_{k} (a_k z_k^2 + b_k z_k^3), \label{eq:chisqnonquadrappr}
\end{equation}
where the coefficients are obtained with
\begin{gather}
	a_k = \frac{\Delta\chi^2}{\delta z^+_k - \delta z^-_k}\left(\frac{\delta z^+_k}{(\delta z^-_k)^2} - \frac{\delta z^-_k}{(\delta z^+_k)^2}\right), \\
	b_k = \frac{\Delta\chi^2}{\delta z^+_k - \delta z^-_k}\left(\frac{1}{(\delta z^+_k)^2} - \frac{1}{(\delta z^-_k)^2}\right).
\end{gather}
This is illustrated in Fig.~\ref{fig:nonquadrappr} (upper diagram), where we show an example of a situation where the $\chi^2$ grows asymmetrically with respect to $z_k$. The quadratic approximation fails to acknowledge this fact and a third order polynomial is needed to reproduce the $\Delta\chi^2$ growth at $\delta z_k^-$ and $\delta z_k^+$. Similarly, as illustrated in Fig.~\ref{fig:nonquadrappr} (lower diagram), the $y_i$ can be expanded in terms of $z_k$ as
\begin{equation}
	y_i(\vec{z}) \approx y_i[S_0] + \sum_k (d_{ik} z_k + e_{ik} z_k^2), \label{eq:ynonquadrappr}
\end{equation}
where
\begin{align}
	\begin{split}
		d_{ik} = \frac{1}{\delta z^+_k - \delta z^-_k}\bigg[&-\frac{\delta z^-_k}{\delta z^+_k}\left(y_i[S_k^+] - y_i[S_0]\right)\\ &+ \frac{\delta z^+_k}{\delta z^-_k}\left(y_i[S_k^-] - y_i[S_0]\right)\bigg],
	\end{split} \\
	\begin{split}
		e_{ik} = \frac{1}{\delta z^+_k - \delta z^-_k}\bigg[&\frac{1}{\delta z^+_k}\left(y_i[S_k^+] - y_i[S_0]\right)\\ &- \frac{1}{\delta z^-_k}\left(y_i[S_k^-] - y_i[S_0]\right)\bigg].
	\end{split}
\end{align}
One should note that the above approximations do not yield a full Taylor expansion to cubic and quadratic order in $\chi^2_\text{old}$ and $y_i(\vec{z})$, respectively, as we have neclected off-diagonal terms proportional to $z_l z_k^2$ and $z_l z_k$ for $l \neq k$. Even so, we will refer to reweighting with these approximations as a \emph{cubic--quadratic} one.

Changing variables to $w_k = 2z_k / (\delta z^+_k - \delta z^-_k)$ and defining $r_k = -\delta z^+_k/\delta z^-_k$, we may alternatively write
\begin{equation}
	\begin{split}
		\chi^2_\text{new}(\vec{w}) - \chi^2_0 \approx &\sum_{k} (A_k w_k^2 + B_k w_k^3)\\ &+ \sum_{ij}\,(y_i(\vec{w}) - y_i^\text{data})\,C^{-1}_{ij}\,(y_j(\vec{w}) - y_j^\text{data}), \label{eq:rechisqnew}
	\end{split}
\end{equation}
where
\begin{gather}
	A_k = \frac{\Delta\chi^2}{4}\left(\frac{1}{r_k^2} + \frac{1}{r_k} + r_k + r_k^2\right), \\
	B_k = \frac{\Delta\chi^2}{8}\left(\frac{1}{r_k^2} + \frac{2}{r_k} - 2\,r_k - r_k^2\right),
\end{gather}
and
\begin{gather}
	y_i(\vec{w}) \approx y_i[S_0] + \sum_k (D_{ik} w_k + E_{ik} w_k^2), \label{eq:reynonquadrappr} \\
	\begin{split}
		D_{ik} = \frac{1}{2}\bigg[&\frac{1}{r_k}\left(y_i[S_k^+] - y_i[S_0]\right)\\ &- r_k\left(y_i[S_k^-] - y_i[S_0]\right)\bigg],  \label{eq:coeffD}
	\end{split} \\
	\begin{split}
		E_{ik} = \frac{1}{4}\bigg[&\left(1 + \frac{1}{r_k}\right)\left(y_i[S_k^+] - y_i[S_0]\right)\\ &+ (1 + r_k)\left(y_i[S_k^-] - y_i[S_0]\right)\bigg]. \label{eq:coeffE}
	\end{split}
\end{gather}

Now, it is a simple numerical task to minimize Equation~\eqref{eq:rechisqnew} with respect to $\vec{w}$. We use MINUIT~\cite{James:1975dr} for the practical applications in the following sections. The found minimum should correspond to that of a full global fit, provided that the approximations~\eqref{eq:rechisqnew} and~\eqref{eq:reynonquadrappr} are good enough. This is not trivially true, but we should expect the approximations work better the closer we are to the original minimum. Thus it makes sense to define a ``penalty term''
\begin{equation}
	P = \sum_{k} (A_k (w_k^\text{min})^2 + B_k (w_k^\text{min})^3) \approx \chi^2_\text{old}(\vec{w}^\text{min}) - \chi^2_0, \label{eq:penalty}
\end{equation}
which essentially counts how much $\chi^2_\text{old}$ has grown from its minimum value, $w_k^\text{min}$ being the values of $w_k$ at the minimum of $\chi^2_\text{new}(\vec{w})$. If $P \ll \Delta\chi^2$, the approximations~\eqref{eq:rechisqnew} and~\eqref{eq:reynonquadrappr} should work well and the reweighted results can be viewed as a proxy for those of a full global fit. Once $P$ grows close to or above $\Delta\chi^2$, the results of reweighting become more sensitive on the made assumptions and one should be cautious on the interpretations. Moreover, a large $P$ signals a tension between the original fit and the new data, which might be due to incompatibilities of some data sets, but can also be caused by an inflexible PDF parametrization, or other limitations of theory description, such as missing higher-order corrections.

The beauty of the reweighting method lies in the fact that the reweighted result for \emph{any} quantity can be obtained simply by using Equation~\eqref{eq:reynonquadrappr}. For example, the new, reweighted, PDFs are obtained by replacing $y_i$ with $f_i$. One should note that while this expression is quadratic in $w_k$, the new PDFs retain a linear dependence on the old ones and thus satisfy the PDF sum rules and DGLAP evolution equations.\footnote{The fact that the new PDFs are linear combinations of the original ones, with a certain weight factor applied to each of them, also justifies the usage of term ``reweighting'' in this context.} This applies also to the new error sets, which can be obtained essentially by following the same procedure as in Section~\ref{sec:hessianuncrt}, with the exception that the new Hessian matrix $\hat{H}_{kl} = \frac{1}{2} \partial^2 \chi^2_\text{new} / \partial w_k \partial w_l|_{\vec{w}=\vec{w}^\text{min}}$ in
\begin{equation}
	\chi^2_\text{new}(\vec{w}) \approx \chi^2_\text{new}(\vec{w}^\text{min}) + \sum_{kl}\,(w_k - w_k^\text{min})\,\hat{H}_{kl}\,(w_l - w_l^\text{min})
\end{equation}
can be put to an explicit form
\begin{equation}
	\begin{split}
		\hat{H}_{kl} =\ &(A_k + 3B_k w_k^\text{min})\delta_{kl}\\ &+ \sum_{ij}(D_{ik} + 2E_{ik} w_k^\text{min})C^{-1}_{ij}(D_{jl} + 2E_{jl} w_l^\text{min})\\ &+ \sum_{ij}(2E_{ik}\delta_{kl})C^{-1}_{ij}(y_j(\vec{w}^\text{min}) - y_j^\text{data}), \label{eq:newhessian}
	\end{split}
\end{equation}
by taking second derivatives of Equation~\eqref{eq:rechisqnew}. Diagonalizing $\hat{H}$ and finding the deviations in the new eigenvector directions corresponding to $\Delta\chi^2$ growth from $\chi^2_\text{new}(\vec{w}^\text{min})$, one obtains the parameter values for the new error sets, using which the uncertainties of any quantity can again be obtained according to Equation~\eqref{eq:reynonquadrappr}.

The cubic--quadratic approximation considered above is not applicable to all cases, as it requires the knowledge of the $\delta z^\pm_k$. Lower-order approximations, initially introduced in Ref.~\cite{Paukkunen:2014zia}, can be obtained from the above results by taking appropriate limits. Taking $r_k \rightarrow 1$ one finds $A_k = \Delta\chi^2$, $B_k = 0$, thus recovering the quadratic approximation for $\chi^2_\text{old}$. In this limit also the definition of the penalty term in Equation~\eqref{eq:penalty} reduces to that of Ref.~\cite{Paukkunen:2014zia}. As $y_i$ retains its quadratic parameter dependence in this limit, we call this a \emph{quadratic--quadratic} approximation. In many cases this is the best option one can resort to, as it only requires access to the PDF error sets and the value of $\Delta\chi^2$. Even simpler, \emph{quadratic--linear}, approximation can be achieved by taking also $E_{ik} \rightarrow 0$. This version is very easy to implement, as finding the new central and error sets in this approximation involves only solving a system of linear equations~\cite{Paukkunen:2014zia}.

\subsection{Comment on non-global tolerances}

The reweighting method can also be extended to non-global tolerances~\cite{Martin:2009iq}, simply by setting
\begin{gather}
	A_k = \frac{1}{4}\left((T_k^+)^2\left(\frac{1}{r_k^2} + \frac{1}{r_k}\right) + (T_k^-)^2\left(r_k + r_k^2\right)\right), \\
	B_k = \frac{1}{8}\left((T_k^+)^2\left(\frac{1}{r_k^2} + \frac{2}{r_k} + 1\right) - (T_k^-)^2\left(1 + 2\,r_k + r_k^2\right)\right),
\end{gather}
where $(T_k^\pm)^2 = \chi^2_\text{old}(\delta z^\pm_k) - \chi^2_0$ are the tolerances of the individual error sets, determined by requiring acceptable values of $\chi^2$ for each individual data set in the original analysis~\cite{Martin:2009iq}. While the new, reweighted central PDF set can be obtained uniquely in this way, the determination of the new error sets involves additional arbitrariness. As the new eigenvector directions obtained by diagonalizing the Hessian matrix in Equation~\eqref{eq:newhessian} are not parallel to the original ones, it is not directly obvious how large tolerances should be allowed in each of these new parameter directions. It was argued in Ref.~\cite{Schmidt:2018hvu} that if the new eigendirections are not significantly rotated away from the original ones, it would be sufficient to use the original tolerances $(T_k^\pm)^2$ also for obtaining the new error sets. While this can work in some cases, it would be advisable to have a measure on the amount of parameter rotations in the reweighting to test whether the limits of this assumption are met. Another possibility would be to use a global tolerance for the reweighted PDFs, e.g.~by taking the average over the $(T_k^\pm)^2$, but this also would lead to changing the error definition from the original one, thus reducing the comparability of the new and old uncertainties. In general, setting the new non-global tolerances reliably would require a complete refit.

\section{CMS 5.02 TeV dijets and their impact on PDFs} \label{sec:dijets}

The CMS dijet data~\cite{Sirunyan:2018qel} consist of distributions of dijet pseudorapidity
\begin{equation}
	\eta_{\rm dijet} = \frac{1}{2}(\eta^{\rm leading} + \eta^{\rm subleading})
\end{equation}
in bins of average transverse momentum of the jet pair
\begin{equation}
	p_{\rm T}^{\rm ave} = \frac{1}{2}(p_{\rm T}^{\rm leading} + p_{\rm T}^{\rm subleading}).
\end{equation}
Here, $\eta^{\rm (sub)leading}$ and $p_{\rm T}^{\rm (sub)leading}$ refer to the pseudorapidity and transverse momentum of the jet with (second to) largest transverse momentum of the event. Jets are defined with the anti-$k_{\rm T}$ algorithm~\cite{Cacciari:2008gp} using a distance parameter $R = 0.3$. The events used in the analysis are required to have a leading jet with transverse momentum $p_{\rm T}^{\rm leading} > 30 \text{~GeV}$ and a subleading jet with $p_{\rm T}^{\rm subleading} > 20 \text{~GeV}$ and the two jets are required to have an azimuthal angle separation $\Delta\phi > 2\pi/3$. In pPb collisions the two jets are required to be in a rapidity interval $-3 < \eta^{\rm lab}_{\rm jet} < 3$ in the laboratory frame. Due to unequal beam energies, $E_{\rm p} = 4 \text{~TeV}$ and $E_{\rm Pb} = \frac{82}{208}E_{\rm p}$, the nucleon--nucleon center-of-mass system is boosted in this frame. To attain corresponding coverages in the center-of-mass frames, CMS measured the pp spectra in the interval $-3.465 < \eta^{\rm lab}_{\rm jet} < 2.535$. Here, as in the CMS publication, the pp data are shifted in pseudorapidity by $+0.465$, so that the measured dijets cover a pseudorapidity range $-3 < \eta_{\rm dijet} < 3$ in both pp and pPb.

The CMS data are self-normalized in each bin of $p_{\rm T}^{\rm ave}$, i.e.\ given in the form
\begin{equation}
	\frac{1}{\mathrm{d}\sigma/\mathrm{d}p_\mathrm{T}^\text{ave}}\,\mathrm{d}^2\sigma/\mathrm{d}p_\mathrm{T}^\text{ave}\mathrm{d}\eta_\text{dijet}.
\end{equation}
This is advantageous due to a partial cancellation of correlated experimental (including luminosity-) uncertainties and theoretical hadronization corrections.\footnote{For a demonstration of cancellation of the hadronization effects in the normalization, see Ref.~\cite{DogaHIJetWorkshop2016}. With the relatively small $R = 0.3$, the contribution from underlying event should be small in the first place.} Accordingly, we do not apply nonperturbative corrections to our predictions. We work at NLO as the NNLO calculations of Ref.~\cite{Currie:2017eqf} are not publicly available at this moment. Our theory calculations are performed with NLOJet++~\cite{Nagy:2003tz} using the anti-$k_{\rm T}$ algorithm through FastJet package~\cite{Cacciari:2011ma}. We fix the factorization and renormalization scales to be the same, $\mu_{\rm F} = \mu_{\rm R} = \mu$, and use $\mu = p_{\rm T}^{\rm ave}$ as our central scale choice to keep consistency with the CT14 and EPPS16 fits, but study also variations around this central scale choice to approximate the magnitude of missing higher-order uncertainties (MHOUs).\footnote{The CT14 analysis uses the individual-jet $p_{\rm T}$ as the scale for the inclusive-jet cross sections. To LO, $p_{\rm T}^{\rm leading} = p_{\rm T}^{\rm subleading}$, and thus using $\mu = p_{\rm T}^{\rm ave}$ for dijets recovers the CT14 scale definition in the $2 \rightarrow 2$ limit.} In all figures, PDF uncertainties are presented with the asymmetric prescription of Equation~\eqref{eq:asympresc}. As the data correlations are not available, we simply add the statistical and systematical uncertainties in quadrature.

\subsection{Proton--proton dijet spectra and CT14 reweighting}

\begin{figure*}
	\centering
	\includegraphics[width=\textwidth]{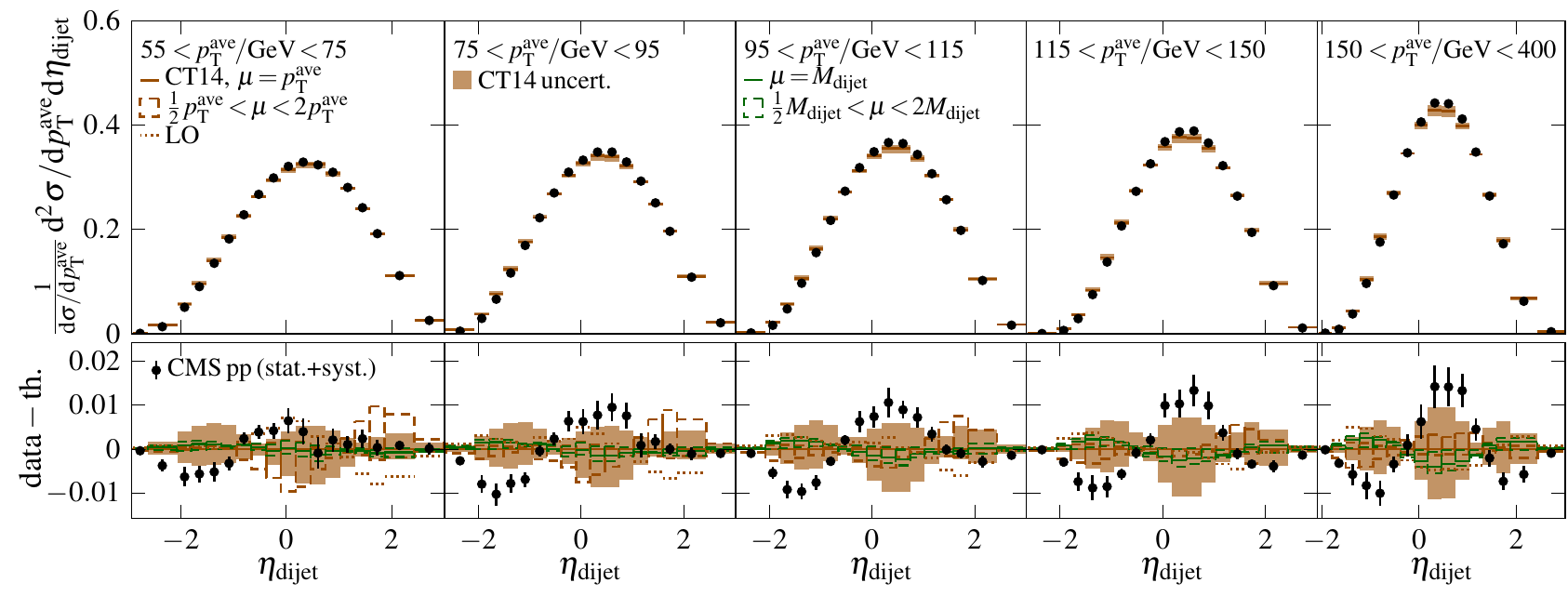}
	\caption{Upper panels: Distributions of dijets in 5.02 TeV proton--proton collisions against $\eta_{\rm dijet}$ and normalized to unity in each bin of $p_{\rm T}^{\rm ave}$. The imposed kinematic cuts are discussed in text. Black markers show the data from the CMS measurement~\cite{Sirunyan:2018qel} with vertical bars showing the statistical and systematical uncertainties added in quadrature. Solid orange lines represent the results from the NLO pQCD calculation using the central set of the CT14 NLO PDFs~\cite{Dulat:2015mca} with $\mu = p_{\rm T}^{\rm ave}$ scale choice, light orange boxes the associated PDF uncertainties from the CT14 NLO error sets. 	Lower panels: Difference to the central CT14 result. Dashed hollow boxes show the dependence of NLO predictions on factor two upward and downward variations of the scale choice. Dotted lines represent the results from the respective LO pQCD calculation. The results with $\mu = M_{\rm dijet}$ scale choice and its factor two variations are indicated in green.}
	\label{fig:ppspectra}
\end{figure*}

The self-normalized pp dijet spectra measured by CMS are shown in Fig.~\ref{fig:ppspectra} along with theory calculations using the CT14 NLO PDFs. While the predictions describe well the $p_{\rm T}^{\rm ave}$ systematics of the data, we see that the predicted pseudorapidity spectra are systematically wider than the measured distributions, with the discrepancy between the data and CT14 central prediction being much larger than the experimental uncertainties, yielding a very poor figure of merit, $\chi^2/N_{\rm data} = 7.5$. To study the possible source of this discrepancy, we show in Fig.~\ref{fig:ppspectra} both the uncertainties from CT14 PDFs, as well as factor of two scale variations around the central scale choice $\mu = p_{\rm T}^{\rm ave}$ and results from a leading order (LO) calculation at the central scale.

We see that in most bins, especially towards high $p_{\rm T}^{\rm ave}$, the discrepancy between the data and theory is larger than the associated scale uncertainty. As the factor of two scale variations often underestimate the true size of higher order corrections (see e.g.~Ref.~\cite{Currie:2017eqf}), not much can be learned from this fact alone. However, as the LO-to-NLO corrections shown in Fig.~\ref{fig:ppspectra} (lower panels) are of the same size as the scale uncertainties, we should not expect the NLO-to-NNLO corrections to be any larger than these. Hence the discrepancy is unlikely to be just due to missing NNLO terms, which in turn points into the direction that the CT14 PDFs need to be modified for a better description of the data. Towards smaller $p_{\rm T}^{\rm ave}$ the scale variation effects become more important, leaving room for improvement with NNLO corrections. Another possible scale choice would be the invariant mass of the dijet, $\mu = M_{\rm dijet}$, a choice which was found in Ref.~\cite{Currie:2017eqf} to yield a better perturbative convergence up to leading-color NNLO precision. We have tested this option, shown also in Fig.~\ref{fig:ppspectra} (lower panels), and report that here at the NLO level it tends to give smaller scale-uncertainty bands especially at low $p_{\rm T}^{\rm ave}$ and that the results do not differ much from the central $\mu = p_{\rm T}^{\rm ave}$ predictions. This points again towards smallness of the NNLO corrections. With even slightly wider predictions, $\mu = M_{\rm dijet}$ gives a worse data description than the $\mu = p_{\rm T}^{\rm ave}$ scale choice, and thus we work with the latter in what follows.

To see the modifications on the CT14 PDFs the CMS dijet data would indicate, we have performed a reweighting study with these data. As most of the data points lie outside the CT14 uncertainties, we could expect the needed modifications to be rather strong. Nominally, the CT14 uncertainties correspond to a global tolerance $\Delta\chi^2 = 100$, but to enforce a $90\%$ confidence level agreement individually with each data set used in the analysis, CT14 uses in addition so called ``Tier-2 penalties''. Hence, the parameter variations $\delta z^\pm_k$ in CT14 do not exactly match with $\pm\sqrt{100}$, but can be somewhat smaller. As no detailed information is available on how large these deviations are, the best we can do is to assume $\chi^2$ to be perfectly quadratic and use $\Delta\chi^2 = 100$. For this reason, we perform the CT14 reweighting in the quadratic--quadratic approximation, noting that the reweighted uncertainties might not be directly comparable with the original ones, and that the new central set underestimates the true impact on CT14, as the use of $\Delta\chi^2 = 100$ overestimates the growth of $\chi^2_\text{old}$ in varying the PDF parameters.

\begin{figure*}
	\centering
	\includegraphics[width=\textwidth]{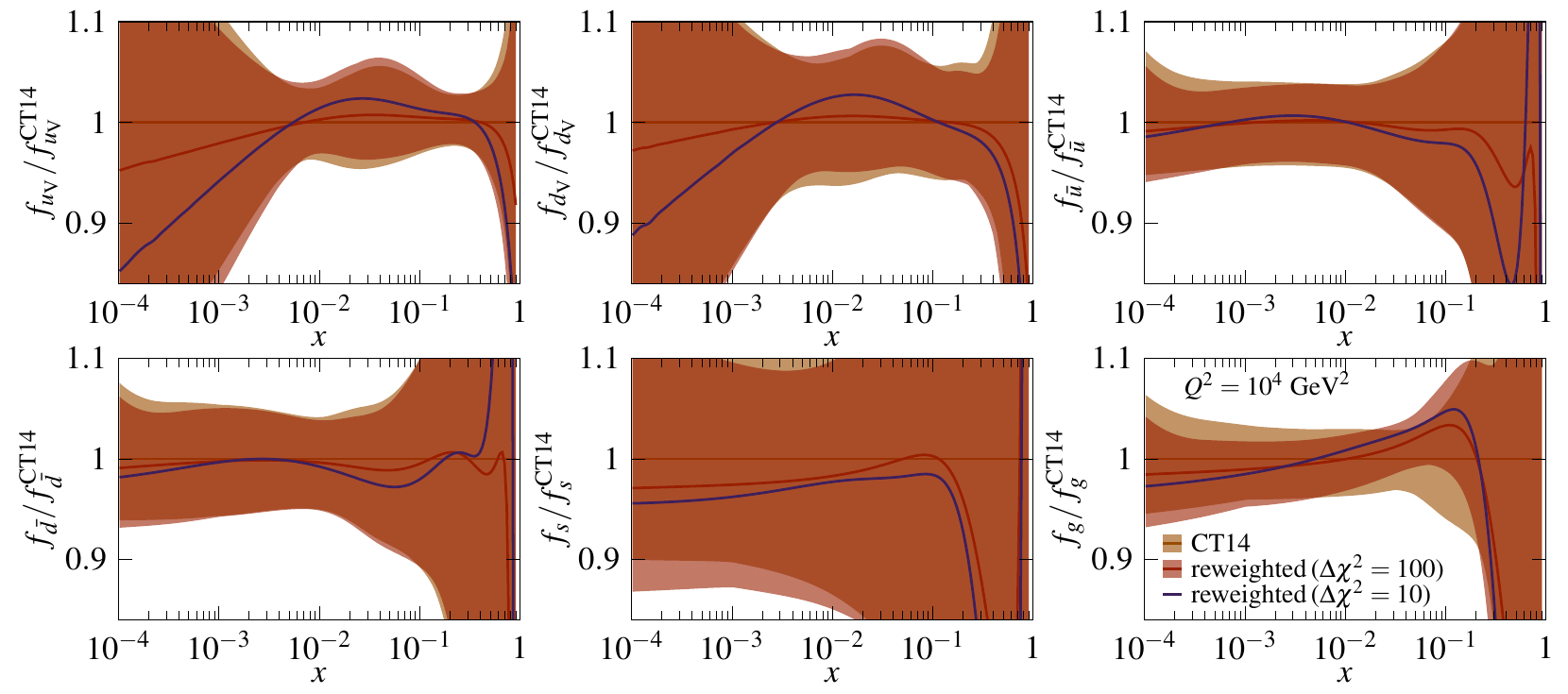}
	\caption{The impact of reweighting on CT14 NLO PDFs at $Q^2 = 10^4~{\rm GeV}^2$. The original CT14 PDFs are shown in orange, with the solid line representing the central set PDFs, the ratio to which is shown in each panel. The corresponding PDFs obtained with quadratic--quadratic reweighting using $\Delta\chi^2 = 100$ are shown in red and the central set of the reweighting with $\Delta\chi^2 = 10$ is presented with a solid purple line.}
	\label{fig:CT14rw}
\end{figure*}

\begin{figure*}
	\centering
	\includegraphics[width=\textwidth]{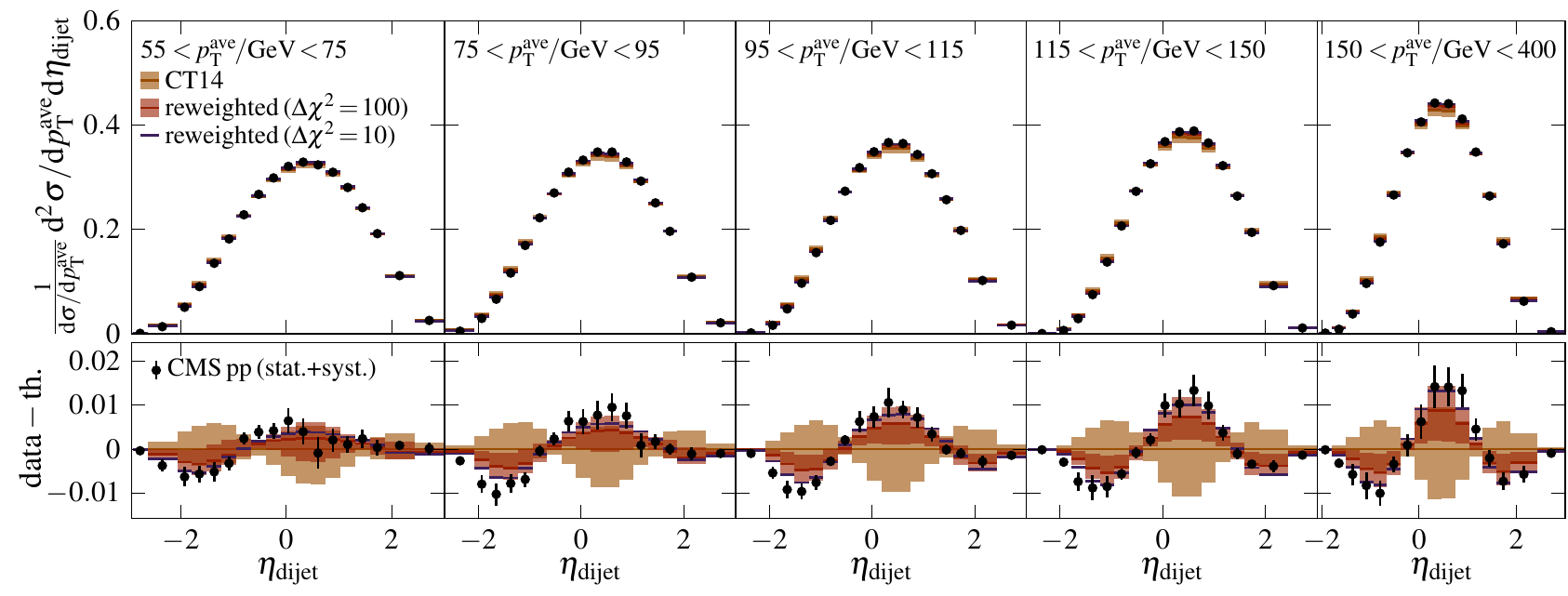}
	\caption{Upper panels: The impact of reweighting on CT14 predictions of pp dijet spectra. The original predictions are shown in orange and the results obtained with quadratic--quadratic reweighting using $\Delta\chi^2 = 100$ are shown in red. In both cases the solid lines corresponding to the central set and the shaded boxes showing the PDF uncertainty. In addition, resulting spectra from reweighting with $\Delta\chi^2 = 10$ are shown as purple lines. Lower panels show again the difference to the original central CT14 results.}
	\label{fig:ppspectrarw}
\end{figure*}

\begin{figure*}
	\centering
	\includegraphics[width=\textwidth]{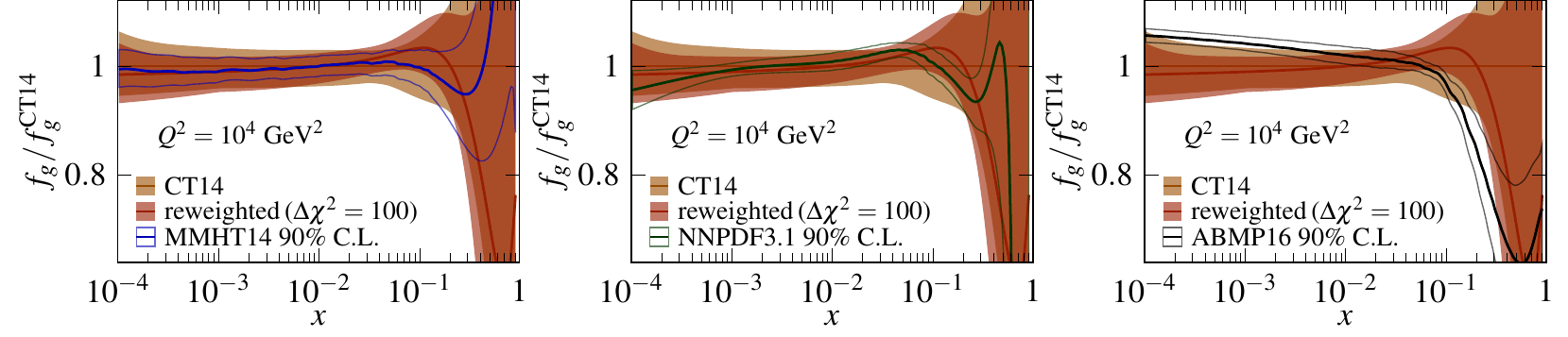}
	\caption{Comparison of the NLO gluon PDFs of the original and reweighted CT14 sets with those from the MMHT14, NNPDF3.1 and 5-flavour ABMP16 analyses. The uncertainty bands of the latter have been scaled with a factor 1.64 to nominally match with the 90\% confidence level definition of the CT14 analysis.}
	\label{fig:CT14pdfcomp}
\end{figure*}

The resulting reweighted PDFs are compared with the original CT14 NLO PDFs in Fig.~\ref{fig:CT14rw}. For all quark flavours, the found modifications are modest compared to the size of PDF uncertainties. Only at very large $x$ we can see a clear downward bend in the central valence-quark PDFs, caused by the fit trying to adapt to the data at large rapidities, where gluon--valence-quark scattering dominates the cross sections. There is a similar, but even more pronounced, large-$x$ depletion for the gluons. In addition, we find an enhancement for gluons at $x \sim 0.1$, compensating for the excess in data at midrapidity. Such modifications to gluon PDF are not totally unexpected. The MMHT14 gluon PDF~\cite{Harland-Lang:2014zoa}, which closely resembles that of CT14, acquires rather similar modifications when confronted with the $7~{\rm TeV}$ high-luminosity inclusive jet data~\cite{Harland-Lang:2017ytb}. Also, attributed to including $8~{\rm TeV}$ differential top-quark data, the NNPDF3.1 fit has large-$x$ gluons suppressed compared to CT14 and MMHT14~\cite{Ball:2017nwa}. In addition, a recent reweighting study using multiple top-production data sets found very similar CT14 modifications as we do here~\cite{Azizi:2018iiq}. Thus, we have evidence that the CT14 gluon distribution is simply too hard to be able to fully describe jet and top-quark measurements.

Fig.~\ref{fig:ppspectrarw} shows the reweighted dijet spectra in comparison to data and original CT14 predictions. The reweighting clearly improves compatibility with the data, especially in the midrapidity region, where the data and theory are now in agreement within the associated uncertainties. At $\eta_{\rm dijet} \lesssim -1$, the data still deviates from the reweighted results. This is also reflected in the figure of merit, $\chi^2/N_{\rm data} = 2.0$, which is still quite high, but vastly better than before the reweighting. For a comparison, we have calculated the dijet spectra also using the MMHT14~\cite{Harland-Lang:2014zoa}, NNPDF3.1~\cite{Ball:2017nwa} and 5-flavour ABMP16~\cite{Alekhin:2018pai} NLO PDFs. These yield $\chi^2/N_{\rm data}$ goodness-of-fit values 4.7, 4.0 and 2.7, respectively, showing that less than perfect agreement with the data is not only a problem with CT14. However, the very strong disagreement between data and CT14 before reweighting appears to be a rather extreme case. In Fig.~\ref{fig:CT14pdfcomp} the gluon PDFs of MMHT14, NNPDF3.1 and ABMP16 are compared with the CT14 before and after the reweighting. The reweighting brings the CT14 gluon distribution to a closer agreement with the other PDFs, particularly at small $x$ to the MMHT14 and NNPDF3.1 and, more importantly, at large $x$ to the NNPDF3.1 and ABMP16. Clearly a reduction in high-$x$ gluons compared to CT14 similar to those in the NNPDF3.1 and ABMP16 fits is preferred by the data.

The penalty term for the reweighted CT14 fit is rather high, with $P/\Delta\chi^2 = 1.17$, clearly indicating that we are reaching the limits of the applicability of the reweighting method. This can be interpreted either as a tension between the dijet data and some datasets used in the CT14 analysis, or as an inflexibility of the CT14 fit form in the high-$x$ region which is probed by the dijets at large rapidities, where the data were not well reproduced and where the data would support even stronger suppression in the PDFs. To test if the CT14 parametrization could adapt to the dijet data, we have performed a reweighting also with an artificially low $\Delta\chi^2 = 10$. In a global fit, this would translate to putting an additional tenfold weight on the new data. The results for the new central PDF set are shown as purple lines in Figures~\ref{fig:CT14rw} and~\ref{fig:ppspectrarw}. With stronger low- and high-$x$ suppression and mid-$x$ enhancement for gluons, this fit achieves a much more reasonable goodness-of-fit $\chi^2/N_{\rm data} = 0.9$ for these data. For this, substantial help from valence quarks, which get strong modifications in this case, is also needed. Still, the data at $\eta_{\rm dijet} \lesssim -1$ are not perfectly reproduced, which might be a signal of a parametrization issue, as the relative contribution from the original fit to the total $\chi^2$ is decreased with the lowered $\Delta\chi^2$. With $P/\Delta\chi^2 = 3.61$, this fit is in a clear tension with the original CT14 analysis. Of course, once the correlations in the dijet data are made available, one should study whether a shift in some of the systematic parameters could improve the fit at $\eta_{\rm dijet} \lesssim -1$. It is also conceivable that the residual disagreement is due to the NNLO corrections.

\begin{figure}
	\centering
	\includegraphics[width=\columnwidth]{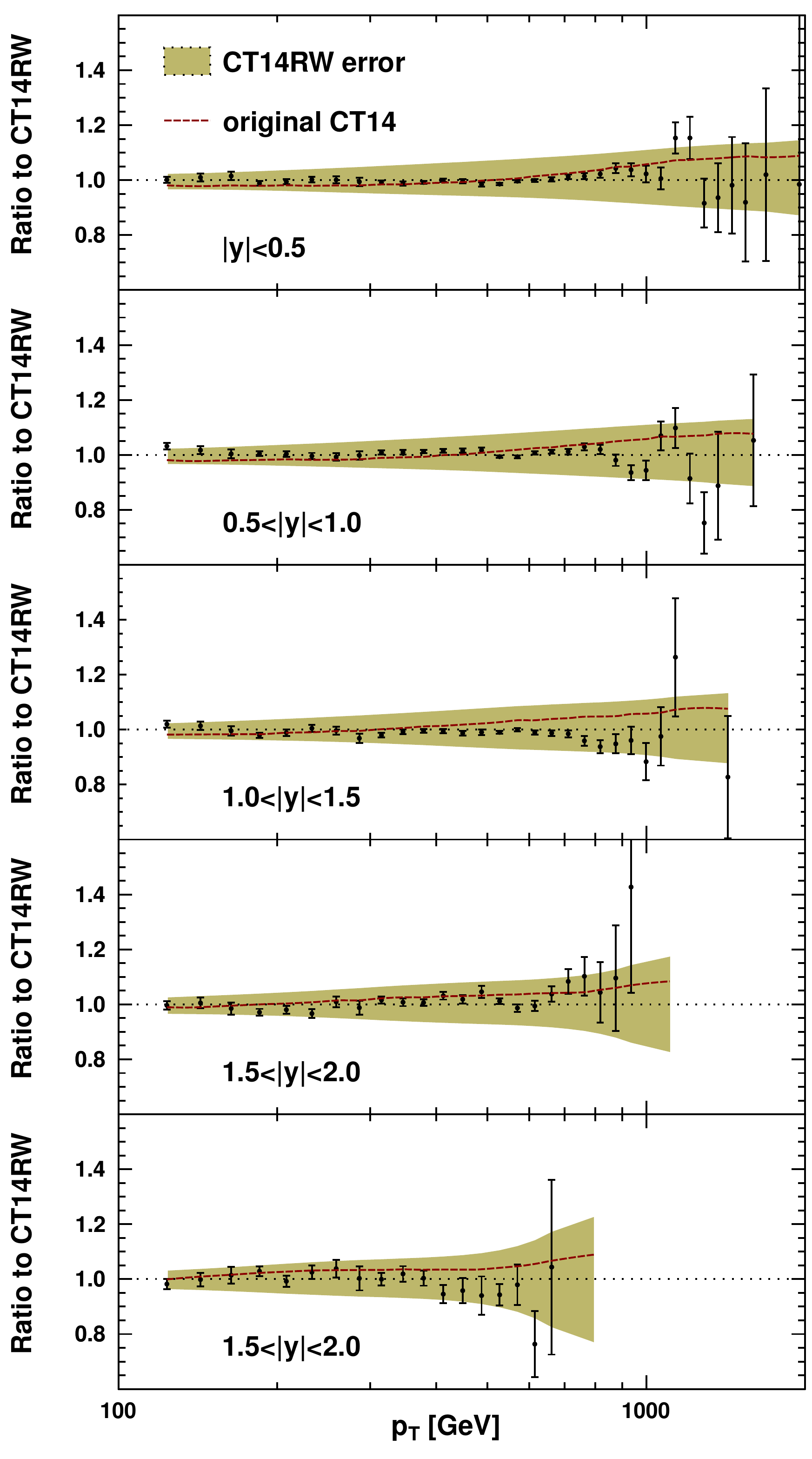}
	\caption{Comparison of CMS 7 TeV inclusive jet measurements~\cite{Chatrchyan:2012bja} and NLO predictions obtained using the CT14 NLO PDFs~\cite{Dulat:2015mca} reweighted with the 5.02 TeV dijet data~\cite{Sirunyan:2018qel}. The optimal systematic shifts in the correlated experimental uncertainties are applied to the data points (similarly as in Ref.~\cite{Paukkunen:2014zia}) and only statistical uncertainties are shown. Dashed red lines show the ratio of predictions with the original CT14 PDFs to those with the reweighted PDFs.}
	\label{fig:ppinclusive}
\end{figure}

\begin{figure*}
	\centering
	\includegraphics[width=\textwidth]{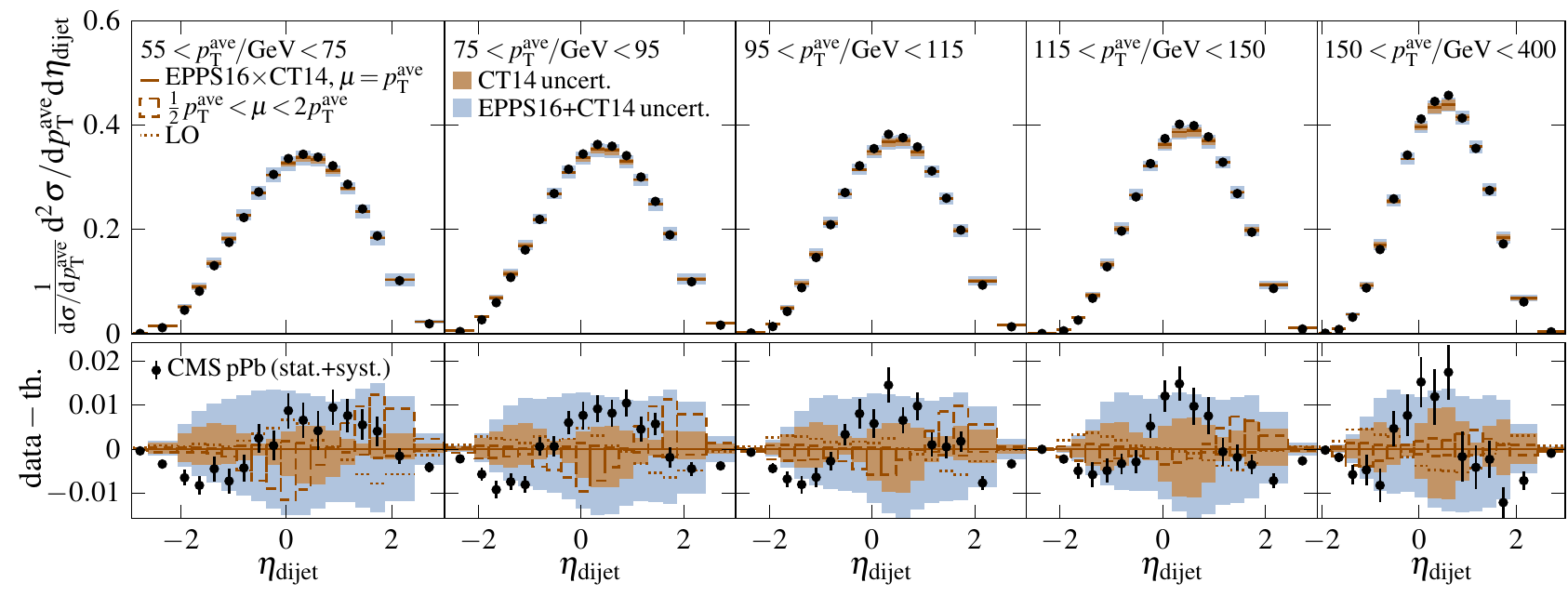}
	\caption{As Fig.~\ref{fig:ppspectra}, but now with pPb data and predictions with EPPS16 nuclear modifications imposed on the CT14 NLO proton PDFs and omitting the results with $\mu = M_{\rm dijet}$ for clarity. Light blue boxes show the combined uncertainty from the CT14 and EPPS16 PDFs.}
	\label{fig:pPbspectra}
\end{figure*}

A comprehensive study of possibly conflicting datasets within CT14 is outside the scope of this article, but as a cross check we have tested the compatibility of the reweighted PDFs with the CMS 7 TeV inclusive jet measurements~\cite{Chatrchyan:2012bja} which are included in the CT14 analysis. For these calculations we use the pre-computed fastNLO grids~\cite{Wobisch:2011ij}, setting the renormalization and factorization scales equal to the transverse momentum $p_{\rm T}$ of the individual jet as in the CT14 analysis. Fig.~\ref{fig:ppinclusive} shows the data-to-theory ratio for the NLO predictions with the CT14 PDFs reweighted with the dijet data using $\Delta\chi^2 = 100$. Also the ratios of the original CT14 central predictions with the reweighted ones are indicated. The data-to-theory agreement happens to be even slightly better for the reweighted PDFs, with $\chi^2/N_{\rm data} = 1.2$, than for the original set, for which $\chi^2/N_{\rm data} = 1.3$. Thus we find that, in the light of reweighting, the CMS measurements of inclusive jets at 7 TeV and dijets at 5.02 TeV are mutually compatible.

\subsection{Significance of proton PDF uncertainties in proton--lead dijet spectra}

The pPb dijet spectra, shown in Fig.~\ref{fig:pPbspectra}, have a rather similar data-to-theory systematics as we had in the pp case. Here, we use the EPPS16 nuclear modifications along with the CT14 NLO proton PDFs in the predictions, i.e. the PDF of a flavour~$i$ in a proton bound in lead at scale~$Q^2$ is obtained with
\begin{equation}
	f^{\rm p/Pb}_i(x,Q^2) = R^{\rm Pb}_i(x,Q^2) f^{\rm p}_i(x,Q^2),
\end{equation}
where $R^{\rm Pb}_i$ is the nuclear modification from the EPPS16 analysis and $f^{\rm p}_i$ the corresponding CT14 PDF of the free proton. The total PDF uncertainties in the cross sections are calculated with
\begin{equation}
	\delta X^\pm_\text{total} = \sqrt{\left(\delta X^\pm_\text{EPPS16}\right)^2 + \left(\delta X^\pm_\text{CT14}\right)^2},
\end{equation}
where $\delta X^\pm_\text{EPPS16}$ are the upward and downward uncertainties obtained with Equation~\eqref{eq:asympresc} using the EPPS16 error sets and keeping the CT14 central set fixed, and $\delta X^\pm_\text{CT14}$, respectively, the uncertainties from the CT14 error sets keeping the EPPS16 central set fixed.

\begin{figure*}
	\centering
	\includegraphics[width=\textwidth]{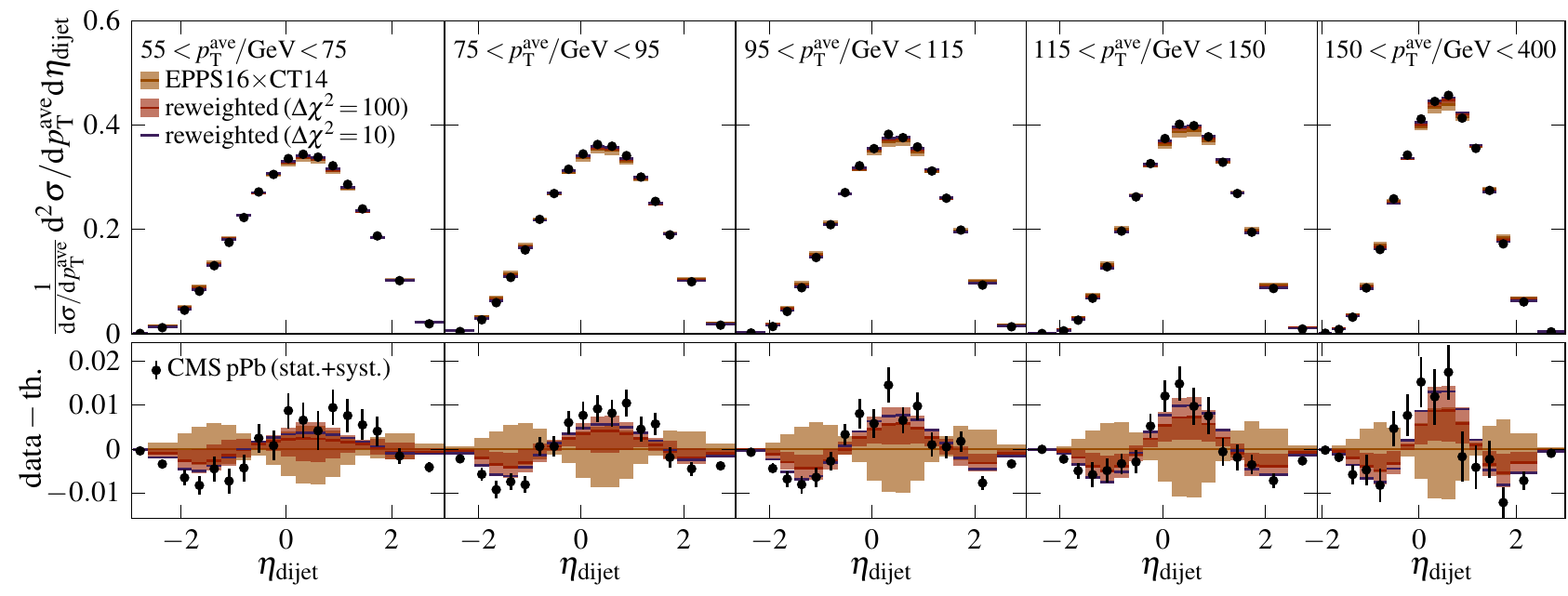}
	\caption{As Fig.~\ref{fig:ppspectrarw}, but now with pPb data and with EPPS16 nuclear modifications imposed on the original and reweighted CT14 PDFs. Only uncertainties from the free-proton PDFs are shown.}
	\label{fig:pPbspectrarw}
\end{figure*}

Again, these predictions give wider distributions than seen in the CMS data, resulting with $\chi^2/N_{\rm data} = 6.9$. While in this case the data points are mostly within the combined nuclear and free-proton PDF uncertainty bands, we can expect that the modifications to the CT14 PDFs, which were found necessary to improve the description of the pp data, play a role also here. Indeed, in Fig.~\ref{fig:pPbspectrarw} we show results with the PDFs obtained by reweighting CT14 with the pp data, observing a clear improvement in the data to theory agreement. We obtain $\chi^2/N_{\rm data} = 2.8$ for the predictions with CT14 reweighted using $\Delta\chi^2 = 100$ and $\chi^2/N_{\rm data} = 1.6$ when using $\Delta\chi^2 = 10$. These numbers are somewhat higher than what we obtained in the pp case, reflecting the fact that also the EPPS16 nuclear modifications need to be adjusted for optimal description of the data. This can also be seen by comparing the data-to-theory agreement in pPb at $\eta_{\rm dijet} \gtrsim 2$ to that in pp: While the CT14 predictions reweighted using $\Delta\chi^2 = 100$ describe well the pp data in these rapidities, the pPb data points lie systematically below the predictions, which hints a preference for deeper nuclear shadowing -- the suppression in the gluon PDF, $R^{\rm Pb}_g < 1$, at small $x$ -- than that in the EPPS16 central set. We will verify this claim in the next section.

An important thing to notice here is that most of the deviations from central theory predictions actually originate from the issues with the free-proton PDFs instead of the nuclear modifications. This large free-proton PDF bias prevents a clean extraction of the PDF nuclear modifications from the pPb spectra. The dijet spectra are certainly not the only pPb observable sensitive to such a free-proton PDF dependence, but the refined proton PDFs found here could also have an effect for example on the predictions for inclusive $t\bar{t}$ production at 8.16 TeV pPb collisions where calculations with CT14+EPPS16 overshoot, but are still compatible with the data~\cite{Sirunyan:2017xku}.

\subsection{Nuclear modification ratio and EPPS16 reweighting} \label{sec:RpA}

\begin{figure*}
	\centering
	\includegraphics[width=\textwidth]{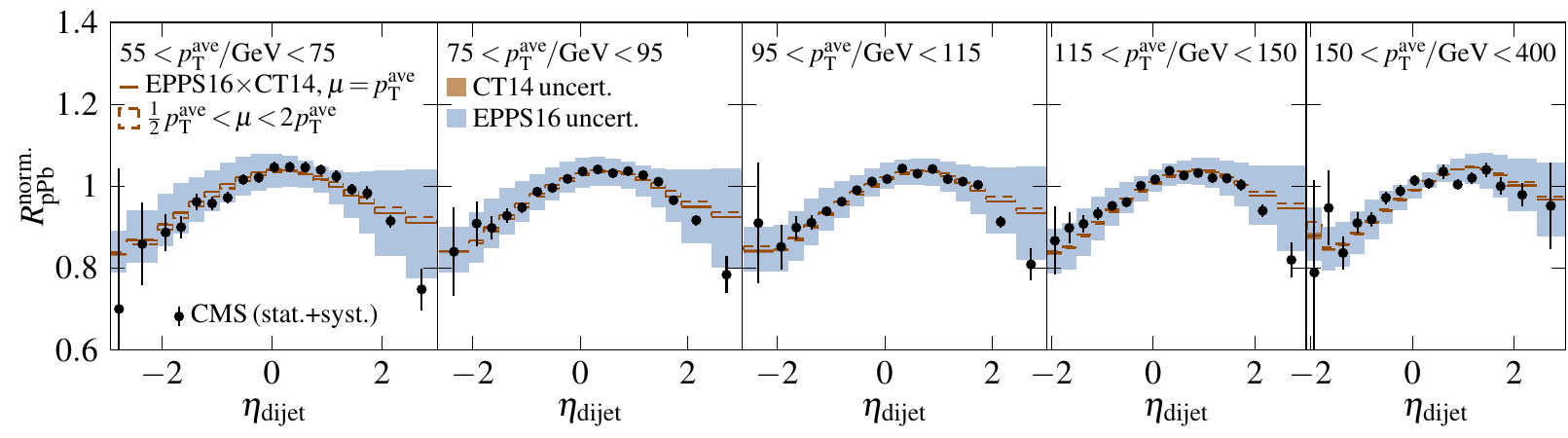}
	\caption{The nuclear modification ratio of normalized pPb and pp differential cross sections. Black markers show the data from CMS measurement~\cite{Sirunyan:2018qel} with vertical bars showing the statistical and systematical uncertainties added in quadrature. Solid orange lines represent the NLO pQCD calculation with $\mu = p_{\rm T}^{\rm ave}$ scale choice using the central set of the CT14 NLO PDFs~\cite{Dulat:2015mca} with EPPS16~\cite{Eskola:2016oht} nuclear modifications.}
	\label{fig:RpPb}
\end{figure*}

\begin{figure*}
	\centering
	\includegraphics[width=\textwidth]{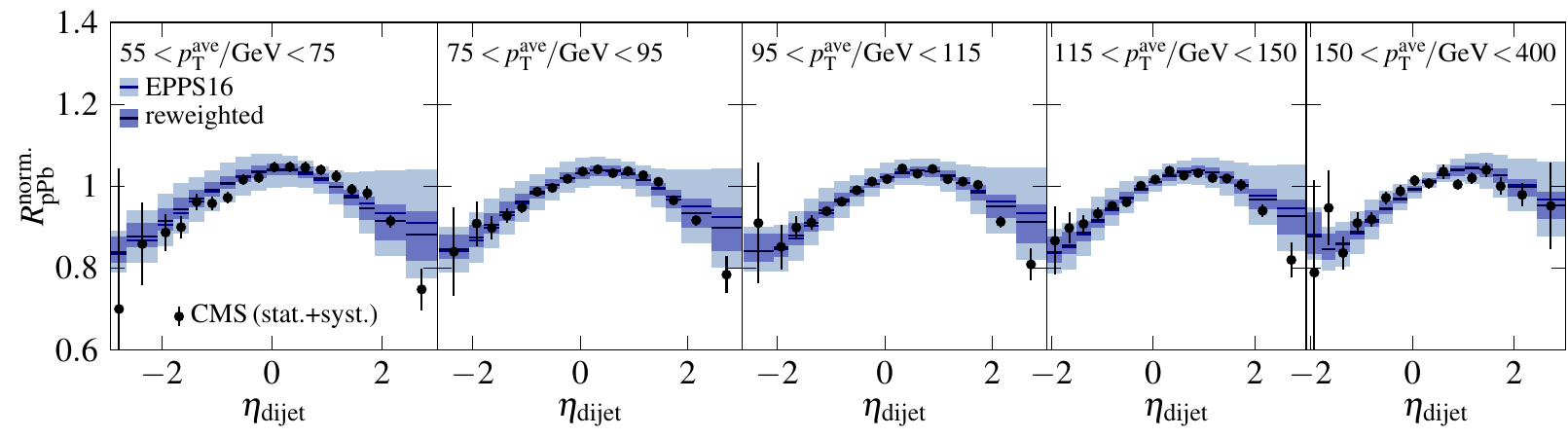}
	\caption{The impact of reweighting on EPPS16 predictions of the nuclear modification ratio of the dijet spectra. The original predictions are shown with solid blue lines and light blue boxes representing the central predictions and the nPDF uncertainties, respectively. The corresponding results after the reweighting are shown with solid black lines and purple boxes.}
	\label{fig:RpPbrw}
\end{figure*}

Let us now consider the nuclear modification ratio of the normalized dijet spectra discussed above, defined as
\begin{equation}
  R_{\rm pPb}^{\rm norm.} = \frac{\frac{1}{\mathrm{d}\sigma^{\rm pPb}/\mathrm{d}p_\mathrm{T}^{\rm ave}}\,\mathrm{d}^2\sigma^{\rm pPb}/\mathrm{d}p_\mathrm{T}^{\rm ave}\mathrm{d}\eta_{\rm dijet}}{\frac{1}{\mathrm{d}\sigma^{\rm pp}/\mathrm{d}p_\mathrm{T}^{\rm ave}}\,\mathrm{d}^2\sigma^{\rm pp}/\mathrm{d}p_\mathrm{T}^{\rm ave}\mathrm{d}\eta_{\rm dijet}}.
\end{equation}
As we have seen that the dijet rapidity distributions in pp and pPb have very similar dependence on the free proton PDFs, we can expect this dependence to efficiently cancel in the ratio. This statement is verified in Fig.~\ref{fig:RpPb}, where we observe the uncertainty band given by CT14 PDFs to be vanishingly small. Also the scale uncertainties, while being larger than the CT14 uncertainties, are small in this observable, implying that MHOUs can be expected to be small as well. This leaves the nuclear modifications as the dominant source of theory uncertainty.

We observe that the CMS data and EPPS16 predictions are in good agreement within the uncertainties. This does not come as a surprise, as part of these data, namely the high-$p_{\rm T}^{\rm ave}$ part of the pPb cross section~\cite{Chatrchyan:2014hqa}, were used in the EPPS16 fit. Still, this agreement is not trivial as with the new pp baseline and being a more differential measurement, these $R_{\rm pPb}^{\rm norm.}$ data contain plenty of new information compared to the 7 data points of forward-to-backward ratios included in the EPPS16 analysis. As was anticipated above, the data points at forward rapidities deviate from the central EPPS16 prediction, indicating a preference for a deeper shadowing in the nPDFs.

Compared to the data, the EPPS16 predictions have much larger uncertainties, which promises a good constraining power when fitting to these data. To study the impact these data would have had in the EPPS16 fit, we have performed a reweighting in the cubic--quadratic approximation introduced in Section~\ref{sec:reweighting}, using $\Delta\chi^2 = 52$ and taking the values of $\delta z^\pm_k$ from Table 2 of Ref.~\cite{Eskola:2016oht}. The results for $R_{\rm pPb}^{\rm norm.}$ are shown in Fig.~\ref{fig:RpPbrw}. Most notably, there is a vast reduction in the EPPS16 uncertainties. Also, at forward rapidities the central prediction comes down a bit, as is expected from the low-lying data points in this region. In the backward direction a slight enhancement in the central prediction can be observed, but this is far less prominent than the suppression in the forward bins. In total, we obtain an improvement in the goodness of fit from $\chi^2/N_{\rm data} = 1.7$ to $1.4$ with a penalty $P/\Delta\chi^2 = 0.14$.

\begin{figure*}
	\centering
	\includegraphics[width=\textwidth]{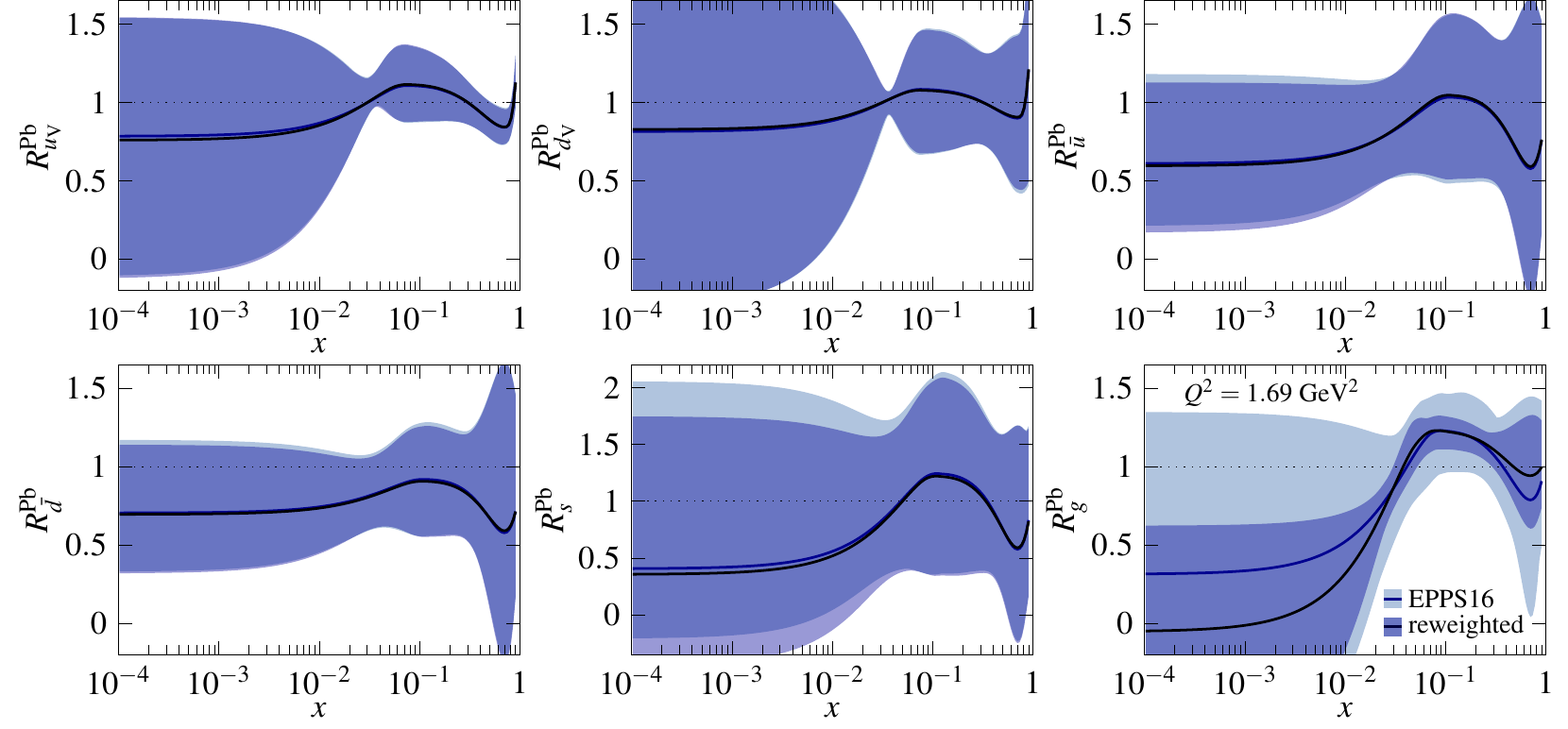}
	\caption{The impact of reweighting the EPPS16 nPDFs with the data on the nuclear modification ratio of the dijet spectra. The original and reweighted EPPS16 nuclear modifications for the lead nucleus are presented at the parametrization scale $Q^2 = 1.69~{\rm GeV}^2$. For better visibility, the $s$-quark modifications are presented with a different vertical axis scaling.}
	\label{fig:EPPS16rw}
\end{figure*}

The corresponding effects on the EPPS16 nuclear modifications in lead at the parametrization scale $Q^2 = 1.69~{\rm GeV}^2$ are presented in Fig.~\ref{fig:EPPS16rw}. There is a striking impact on gluon modification uncertainties, which are reduced across all $x$. In the best-constrained mid-$x$ region, the uncertainties are reduced to less than half of their original size. As the uncertainty band lies clearly above unity in this region, we find strong evidence for gluon antishadowing in lead. At small $x$, the reweighted uncertainty band goes respectively below unity, giving evidence for gluon shadowing. These findings are in accordance with those of Ref.~\cite{Kusina:2017gkz}, where inclusive heavy-flavour production data from measurements at the LHC were used to study the gluon PDF modifications in nuclei. As expected from inspecting the ratio of the dijet spectra, the new central set seems to support stronger shadowing than in the original EPPS16 central fit.

\begin{figure*}
	\sidecaption
	\includegraphics[width=0.673\textwidth]{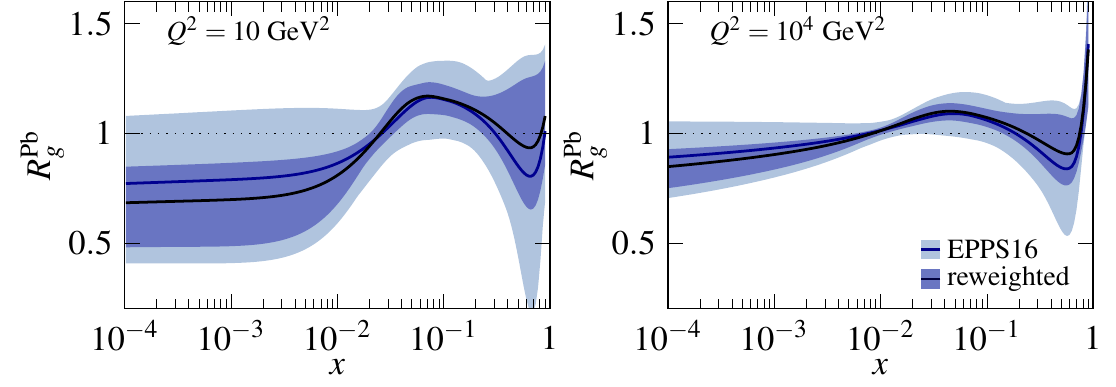}
	\caption{The EPPS16 gluon nuclear modifications in Pb at the scales $Q^2 = 10~{\rm GeV}^2$ and $Q^2 = 10^4~{\rm GeV}^2$ before and after reweighting with the dijet data.}
	\label{fig:EPPS16rwhighscale}
\end{figure*}

Even with the increased gluon shadowing, the most forward bins of $R_{\rm pPb}^{\rm norm.}$ are not well reproduced by the reweighted results, which is also the reason why the $\chi^2/N_{\rm data}$ remained somewhat high even after the reweighting. To be consistent with these forward data points, a very deep shadowing for the gluons would be required. Moreover, the probed $x$ region changes very little between the last and second-to-last $\eta_{\rm dijet}$ data point, and thus such a steep drop as that suggested by the data is difficult to attain. This is because the DGLAP evolution efficiently smooths out even steep structures in the gluon nuclear modification, as can be seen in Fig.~\ref{fig:EPPS16rwhighscale} where we show the gluon nuclear modifications evolved to higher scales. We also note that the systematic uncertainty dominates in the last $\eta_{\rm dijet}$ bins, and thus taking into account the data correlations, once available, could improve the fit quality. These findings should, in the future, be contrasted also with the recent ATLAS conditional yield measurement, where an order of $10$--$20\%$ nuclear suppression for dijets was found in the most forward configuration~\cite{Aaboud:2019oop}.

Also at large $x$, the reweighted gluon modifications are better constrained than in the original EPPS16 analysis. The new central set has $R_g^{\rm Pb}$ closer to unity at $x$ around $0.7$. This is partly enforced by momentum sum rule in combination with the stiffness of the EPPS16 fit function and the deepened small-$x$ shadowing. In any case, the uncertainty remains large, and either an enhancement or a suppression for gluons is possible in this region. On this basis, the conclusion made in Ref.~\cite{Sirunyan:2018qel}, that the dijet data would give evidence of large-$x$ gluon suppression, seems premature. This claim was based on comparison of the data with EPS09~\cite{Eskola:2009uj} and DSSZ~\cite{deFlorian:2011fp} nPDFs, where the former, with gluon suppression at large $x$, agreed well with the data at backward rapidities, but the latter, having the nuclear gluons unmodified, did not. However, going towards backward rapidities, and thus larger $x$ from the Pb side, the contribution of nuclear quarks to the dijet cross section grows rapidly. Hence the difference in predictions with EPS09 and DSSZ in this region has a large contribution from different valence quark modifications. As DSSZ has much smaller large-$x$ suppression for valence quarks than EPS09 (see e.g.\ Ref.~\cite{Paukkunen:2014nqa}), this also partly explains the difference in the dijet predictions of Ref.~\cite{Sirunyan:2018qel}.

\begin{figure*}
	\centering
	\includegraphics[width=\textwidth]{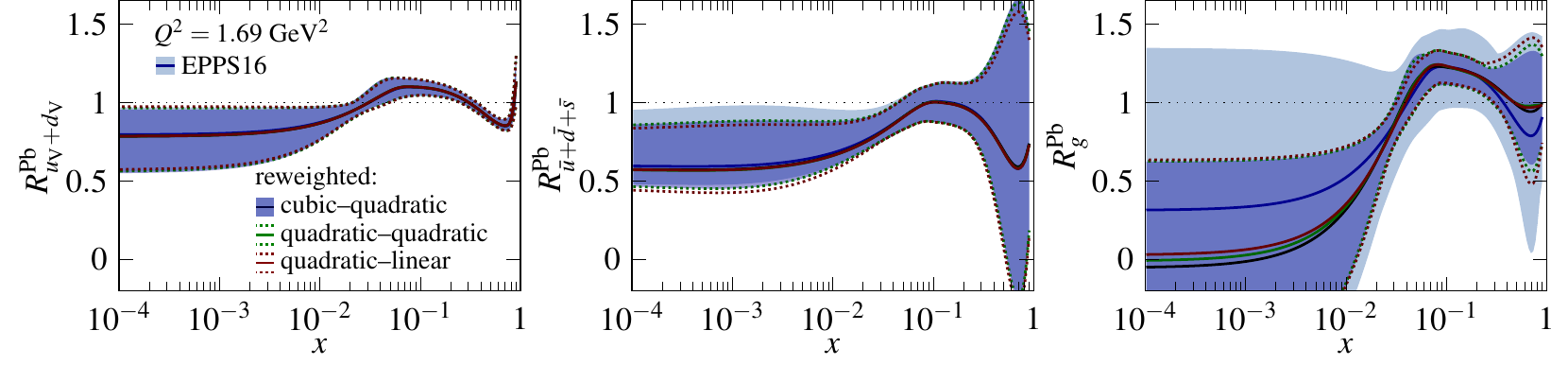}
	\caption{The impact on the average valence and sea quark and gluon modifications under different approximations in the reweighting.}
	\label{fig:EPPS16rwquadlin}
\end{figure*}

On these grounds, it might appear surprising that the dijet data are not able to constrain the valence quark modifications at all, as can be seen from the first two panels in Fig.~\ref{fig:EPPS16rw}. The reason for this is that due to smallness of isospin corrections~\cite{Eskola:2013aya}, the backward dijet data mainly probe the average valence modifications,
\begin{equation}
	R_{u_{\rm V} + d_{\rm V}}^{\rm Pb} = \frac{u^{\rm p/Pb}_{\rm V}+d^{\rm p/Pb}_{\rm V}}{u^{\rm p}_{\rm V}+d^{\rm p}_{\rm V}},
\end{equation}
shown in Fig.~\ref{fig:EPPS16rwquadlin}. This combination is much better constrained than the individual flavours shown in Fig.~\ref{fig:EPPS16rw} and has vastly smaller uncertainties at large $x$ than the gluon modifications. Thus, while large-$x$ valence quarks dominate the dijet cross section at backward rapidities, the uncertainty in the EPPS16 predictions in this region comes dominantly from the less-constrained gluons, and hence it is the gluon modifications which are constrained in the reweighting. Fig.~\ref{fig:EPPS16rwquadlin} shows also the average sea quark modification
\begin{equation}
	R_{\overline{u}+\overline{d}+\overline{s}}^{\rm Pb} = \frac{\overline{u}^{\rm p/Pb}+\overline{d}^{\rm p/Pb}+\overline{s}^{\rm p/Pb}}{\overline{u}^{\rm p}+\overline{d}^{\rm p}+\overline{s}^{\rm p}},
\end{equation}
which is the dominant quark combination constrained at forward rapidities. We observe a modest reduction in the small-$x$ uncertainty, much smaller than that for the gluons. At the level of individual flavours, shown in Fig.~\ref{fig:EPPS16rw}, these constraints affect mostly the $s$-quark modifications, which were poorly constrained in EPPS16.

\subsection{Importance of non-quadratic and non-linear terms in reweighting}

We may now ask whether the inclusion of higher-order (non-quadratic and non-linear) components in the reweighting had a sizable effect on our results. Fig.~\ref{fig:EPPS16rwquadlin} shows the impact of the dijet data on the EPPS16 nuclear modifications in all three approximations discussed in Section~\ref{sec:reweighting}. While, for simplicity of presentation, we show only the average valence and light-sea-quark modifications in addition to those for gluons, the conclusions below apply to individual flavours as well. We find that the cubic--quadratic and quadratic--quadratic approximations give almost identical results. This is rather easy to understand: The new data are precise enough to dominate the shape of the total $\chi^2$ function in the parameter directions that it constrains (mainly those related to gluon degrees of freedom), making the non-quadratic components sub-dominant in the reweighting. Moreover,  as the new central set does not divert far from the original, we are working in a region where the quadratic approximation for $\chi^2_\text{old}$ is rather good. Under different circumstances this might not be the case and the cubic terms could alter the reweighting results significantly.

Next, we consider the reweighting results in the quadratic--linear approximation. Here, we use the linear approximation for the cross sections, but decide to keep the quadratic dependence in the PDFs for better comparability.\footnote{Note that using a linear parameter dependence for the PDFs would render the PDF uncertainties to be perfectly symmetric, so that the comparison with cubic--quadratic and quadratic--quadratic approximations would be meaningful only under the symmetric prescription of Equation~\eqref{eq:sympresc}.} Again, the differences to the results of the cubic--quadratic approximation are rather modest, though for the high-$x$ gluons the quadratic--linear approximation seems to suggest slightly less stringent constraints. The similarity of results in the different approximations can also be seen as a reassuring fact: the results of reweighting do not seem to depend on minute details of our method and we seem to be able to make reliable conclusions based on rather limited information about the original global analysis, at least in this particular case. The obtained results are thus not likely to change if even higher-order contributions are added.

\section{Summary and conclusions}

In this work, we have presented a non-quadratic extension of the Hessian PDF reweighting introduced in Ref.~\cite{Paukkunen:2014zia} and applied the method in the context of CMS dijet measurements at 5.02 TeV. This improved method makes use of the knowledge of parameter variations at which the error sets of the original PDFs are defined, to solve for cubic components of the $\chi^2$ function before inclusion of new data. Similarly, quadratic components in the responses of observables to parameter variations were taken into account. The additional information needed in this cubic--quadratic approximation prevented us from using it when reweighting the CT14 NLO PDFs with the pp dijet distributions, where we had to resort to a simpler quadratic--quadratic approximation, but we were able to apply it to reweight the EPPS16 nPDFs, for which the needed information is available, with the nuclear modification ratio of the dijet spectra. While no large differences were found in the results of reweighting EPPS16 in the cubic--quadratic or quadratic--quadratic approximation, this observation was limited to one specific case, and under different circumstances the cubic terms could become more important. We thus encourage PDF fitters to publish the details of their analysis to a sufficient accuracy, such that the reweighting including the higher-order terms becomes possible. This can be done by publishing the numerical values of the $\delta z^\pm_k$ parameters as defined in Section~\ref{sec:hessianrw} in addition to the tolerance $\Delta\chi^2$. Care must be taken in communicating which error set corresponds to each of these values, so that there is no chance of misinterpretation e.g.\ in what is called a ``plus'' and what a ``minus'' direction. A neat way to do this with LHAPDF~\cite{Buckley:2014ana} would be to set in each PDF grid file a custom flag such as ``{\tt ParamVal}'' to hold the value $\delta z^\pm_k$. These parameter values could then be retrieved by using the method {\tt info().get\_entry("ParamVal")} for each of the PDF error sets.

Comparing the measured pp dijet pseudorapidity spectra with theory calculations using the CT14 NLO PDFs revealed a large discrepancy. We showed that at high $p_{\rm T}^{\rm ave}$ this difference is larger than the associated scale uncertainties and exceeds the size of the NLO corrections, thus being unlikely due to missing NNLO terms alone. This suggested the need for modifying the CT14 PDFs to reach a better agreement with the data. In reweighting CT14 with the dijet data, the gluon PDF acquired significant modifications, especially at large $x$, where a substantial reduction was observed. We discussed also evidence from other studies pointing into the same direction. After reweighting, a much more reasonable $\chi^2$ value for the dijet data was found, but this came with a price of a rather high penalty term, i.e.\ the new central set had diverted quite far from the original minimum. The reason for this apparent discrepancy between CT14 and the dijet data remains elusive. We tested the reweighted PDFs against CMS 7 TeV inclusive jet measurements finding good agreement, and thus no conflict between the considered dijet and inclusive jet data. By performing a reweighting with an artificially low $\Delta\chi^2$, we showed that the CT14 PDFs still had trouble in reproducing the data at $\eta_{\rm dijet} \lesssim -1$, signaling a possible parametrization issue, although NNLO corrections and correlated systematics can also play a role here. Solving this issue is beyond the reach of the reweighting tools and should be studied in the context of a global analysis.

Similar discrepancy as seen with the pp spectra is observed also in the case of pPb. We showed that applying the same CT14 modifications as found in the reweighting with pp data substantially improves the data-to-theory agreement also in pPb. As the pPb dijet distributions contain a substantial free-proton PDF dependence, a clean extraction of their nuclear modifications is not possible from these data directly. Taking the ratio of the pPb and pp spectra, however, leads to a very efficient cancellation of not only the free-proton uncertainties but also of the scale uncertainties, thus giving an excellent probe of the nPDFs. We showed that the measured nuclear-modification ratio of dijet spectra is in a good agreement with the NLO predictions using the EPPS16 nPDFs. Some deviation from the EPPS16 central prediction was observed at $\eta_{\rm dijet} \gtrsim 2$, supporting a stronger shadowing for gluons than present in the EPPS16 central set. As a whole, these data give compelling evidence of small-$x$ gluon nuclear shadowing and mid-$x$ antishadowing, as was revealed in reweighting the EPPS16 nPDFs. We obtained significant new constraints on the EPPS16 gluon modifications in lead throughout the probed range, reducing the uncertainties even to less than half of their original size.

\begin{acknowledgements}
We thank Yen-Jie Lee for discussions. We have received funding from the Academy of Finland, Project 297058 of K.J.E.\ and 308301 of H.P.; P.P.\ acknowledges the financial support from the Magnus Ehrnrooth Foundation. We thank the Finnish IT Center for Science (CSC) for the computational resources allocated under the Project jyy2580.
\end{acknowledgements}

\end{document}